# Data Encryption based on 9D Complex Chaotic System with Quaternion for Smart Grid

Fangfang Zhang, Zhe Huang, Lei Kou*, Yang Li, *Senior Member*, *IEEE*,

Maoyong Cao, Fengying Ma

*Abstract*—With the development of smart grid, the operation and control of power system is realized through power communication network, especially the power production and enterprise management business involve a large amount of sensitive information, and the requirements for data security and real-time transmission are gradually improved. In this paper, a new 9D complex chaotic system with quaternion is proposed for the encryption of smart grid data. Firstly, a new 9D complex chaotic system with quaternion is proposed, and its attractors, bifurcation diagram, complexity, and 0-1 test are analyzed. Secondly, the pseudo-random sequences are generated by the new chaotic system to encrypt power data. Finally, the proposed encryption algorithm is verified with power data and images in the smart grid, which can ensure the encryption security and real-time. The verification results show that the proposed encryption scheme is technically feasible and available for power data and image encryption in smart grid.

## I. INTRODUCTION

Because the power system requires real-time transmission of data communication, most security applications in the computer network can not be implemented in the power system [1]. In the smart grid, most of the critical real-time data is transmitted on the network in plain text, and real-time data encryption is not involved in real-time control. The important data is still in the form of plain text, which poses major safety hazards to the safe and stable operation of the power system [2], [3]. The safe operation of power equipment is directly effected by the function of remote adjusting, remote control and remote measure in the smart grid. Especially when the power information network is maliciously attacked, it results in data loss, tampering and seriously threatens the normal operation of the power information network and then affects the power physical network [4]. Finally, the malicious attack causes a chain reaction in the power network. In extreme cases, the malicious attack propagate alternately between the power information network and power physical network. Therefore, the safe and stable operation of the power grid can not be guaranteed. For example, once the power information network is attacked and the monitor data is tampered with abnormal data. It may cause the power equipment to shut down and bring huge economic losses. For purpose of preventing malicious attackers from sending false abnormal monitor data, it is inevitable to prevent data leakage.

In addition, when encountering the suspected abnormal monitor data, it is crucial to confirm the condition of the equipment as soon as possible [5]. And the maintenance personnel should take photos of the equipment and send them to the control center. Timely reporting of the power equipment status can effectively improve the operation efficiency of the smart grid. Due to data information security, the report is forced to delay. Therefore, it is of great practical significance to study an encryption algorithm that takes into account both confidentiality and real-time.

A deterministic aperiodic model was proposed by Lorenz who is a meteorologist at MIT. He was convinced that chaos is a science that uses fractal geometry to analyze and study the nonlinear dynamics shown in phenomenons such as the butterfly effect. Grassber designed a method to reconstruct the dynamic system. It makes the chao enter the practical application stage by introducing the Lyapunov exponent. By the end of the 20th century, chaos theory began to integrate with other subjects, such as chaotic secure communication [6]-[9], chaotic cryptography [10], [11], chaotic economics and so on. The chaotic signal, which has inherent randomness, ergodicity and sensitivity to initial conditions, is generated by a deterministic system. These characteristics are very similar to the diffusion and scrambling properties in Shannon classical

This work is supported by International Collaborative Research Project of Qilu University of Technology (No.QLUTGJHZ2018020), National Nature Science Foundation of China (No.61773010 and No.61903207), Major scientific and technological innovation projects of Shandong Province (No.2019JZZY010731 and No.2020CXGC010901).

Fangfang Zhang, Zhe Huang, Maoyong Cao, and Fengying Ma are with School of Electrical Engineering & Automation, Qilu University of Technology (Shandong Academy of Sciences), Jinan 250353, China (e-mail: zhff4u@163.com; H794607953@163.com; cmy@qlu.edu.cn; mafengy@163.com).

Lei Kou, is with the Institute of Oceanographic Instrumentation, Qilu University of Technology (Shandong Academy of Sciences), Qingdao 266000, China (Corresponding author. e-mail: koulei1991@hotmail.com).

Yang Li is with the School of Electrical Engineering, Northeast Electric Power University, Jilin 132000 , China (e-mail: liyang@neepu.edu.cn).

model. Therefore, chaos theory has a wide application prospect in the field of information security.

And the dynamic behavior of high dimensional complex chaotic systems are more complicated than real chaotic systems. The real part and the imaginary part are independent in complex-valued chaotic systems, which improves the ergodicity of chaotic system. Hence, high dimensional complex chaotic systems can overcome the disadvantages of short period, uneven distribution of chaotic sequences and small key space. These belong to the low-dimensional chaotic system.

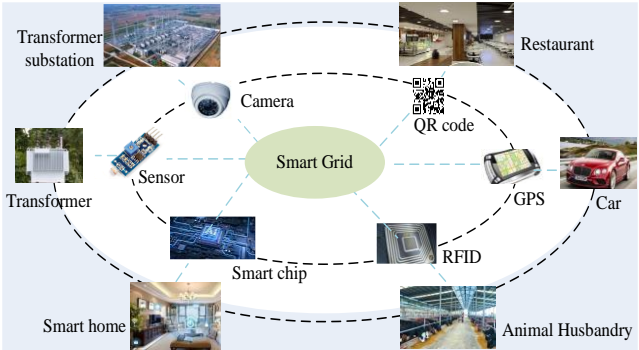

Fig. 1.  The Smart Grid Applications

The basic mathematical concept of quaternion [12]-[14] was proposed by the famous mathematician Hamilton in 1843. He introduced a function called the complex plane function and got a four-dimensional space. The quaternion can be regarded as an extension of a simple complex number. The plural number is mainly composed of a real part and an international unit i as an imaginary number. In the same way, quaternion can also be regarded as composed of a real part and three imaginary units i, j and k.

Many researchers have developed a great number of encryption schemes for the smart grid. In February 2016, Liu *et al.* [15] proposed a lightweight authenticated communication scheme for the smart grid, which ensures a secure two-way communication between the smart meters and the neighborhood gateway. In June 2017, Saxena and Grijalva [16] proposed a novel scheme based on dynamic secrets and encryption with secret keys. The scheme generates a series of dynamic secrets over the communication network, which are used to generate secret keys for data encryption. In June 2019, Gope and Sikdar [17] proposed a lightweight and privacy-friendly masking-based spatial data aggregation scheme for secure forecasting of power demand in smart grids. In March 2020, Song *et al.*[18] designed a dynamic membership data aggregation scheme by the homomorphic encryption and ID-based signature. It reduces the complexity on a new user's joining and an old user's quitting. In May 2021, Qian *et al.* [19] propose a lightweight t-times homomorphic encryption scheme, which can reduce the computational cost of smart devices further and resist quantum attacks.

Although scholars have proposed good encryption schemes for smart grid, there still exist two problems to be solved:

(1) There is no discussion on image transmission encryption for smart grid. Compared with traditional text information, digital image contains more data.

(2)The encryption and decryption time are not discussed.

In order to overcome the above shortcomings, a new complex chaotic system with quaternion is proposed in this paper. This new chaotic system is 9-dimensional and has good encryption performances. The novelty and contributions of this paper are summarized as follows.

(1)A novel complex chaotic system with quaternion is proposed. And the dynamic characteristics are discussed. At the same time, there are few studies on the chaotic system with quaternion.

(2)In order to ensure real-time transmission in the smart grid, an encryption algorithm based on the novel chaotic system is proposed. And the image and the check code, which are encryted by the encryption algorithm, can be sent in time.

(3)Besides, it also can protect information from being tampered. The control center can identify the correct information and will not be misled to turn off the equipment for no reason. Therefore, economic losses are avoided.

The specific content is arranged as follows: The dynamics characteristics of the proposed chaotic system are discussed in Section II. An encryption algorithm is expounded and compared with other algorithms by the usage of some security performances in Section III. In Section IV, the proposed encryption scheme is verified with data and images in the smart grid, and the security analysis is provided. The main work of this paper is summarized in Section V.

## II. 9D Complex Chaotic System with Quaternion

According to [20], the mathematical expression of complex Chen chaotic system is

$$
\begin{cases}
\dot{x}_1 = a(x_2 - x_1) \\
\dot{x}_2 = (b-a)x_1 + bx_2 - x_1 x_3 \\
\dot{x}_3 = (\overline{x}_1 x_2 + x_1 \overline{x}_2)/2 - cx_3
\end{cases} \tag{1}
$$

where $a, b$ are positive constants. $x_1, x_2, x_3$ are independent variables.

Then the mathematical expression of the quaternion is defined by

$$
\begin{cases}
x_1 = u_1 + iu_2 + ju_3 + ku_4 \\
x_2 = u_5 + iu_6 + ju_7 + ku_8 \\
x_3 = u_9
\end{cases} \tag{2}
$$

where $u_i (i=1,2,\ldots,9)$ are independent variables. $i, j, k$ are imaginary units.

System (1) is extended to the quaternion field. And bring (2) into (1). Finally separate the real part from the imaginary part and obtain:

$$\begin{cases} \dot{u}_1 = a(u_5 - u_1) \\ \dot{u}_2 = a(u_6 - u_2) \\ \dot{u}_3 = a(u_7 - u_3) \\ \dot{u}_4 = a(u_8 - u_4) \\ \dot{u}_5 = (b-a)u_1 + bu_5 - u_1 u_9 \\ \dot{u}_6 = (b-a)u_2 + bu_6 - u_2 u_9 \\ \dot{u}_7 = (b-a)u_3 + bu_7 - u_3 u_9 \\ \dot{u}_8 = (b-a)u_4 + bu_8 - u_4 u_9 \\ \dot{u}_9 = u_1 u_5 + u_2 u_6 + u_3 u_7 - c u_9 \end{cases} \quad (3)$$

where $a, b, c$ are positive constants.

Next, the dynamic characteristics of the system are intuitively analyzed from the chao attractor, bifurcation diagram, 0-1 test, and complexity analysis.

### A. Chaos Attractor

Set $a = 27$, $b = 23$, $c = 1$. For initial conditions (0.1, 0.1, 0.1, 0.1, 0.1, 0.1, 0.1, 0.1, 0.1), the phase portraits of attractor are depicted in Fig. 2.

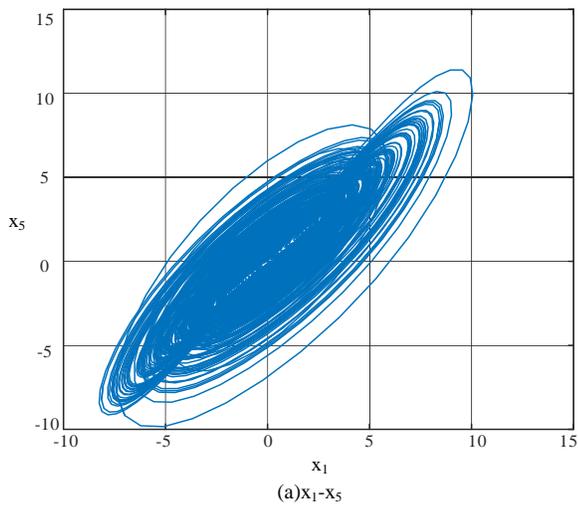

(a)$x_1$-$x_5$

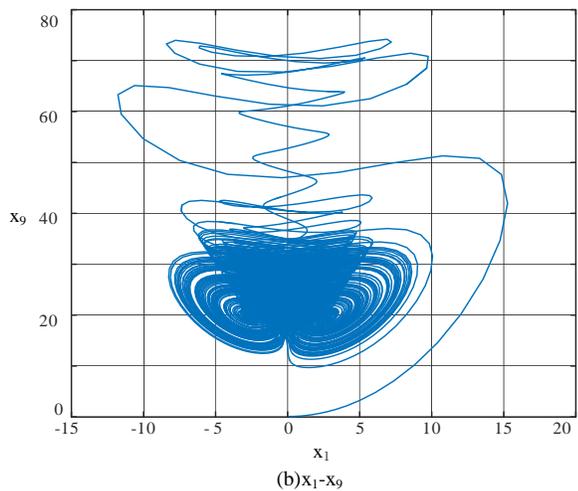

(b)$x_1$-$x_9$

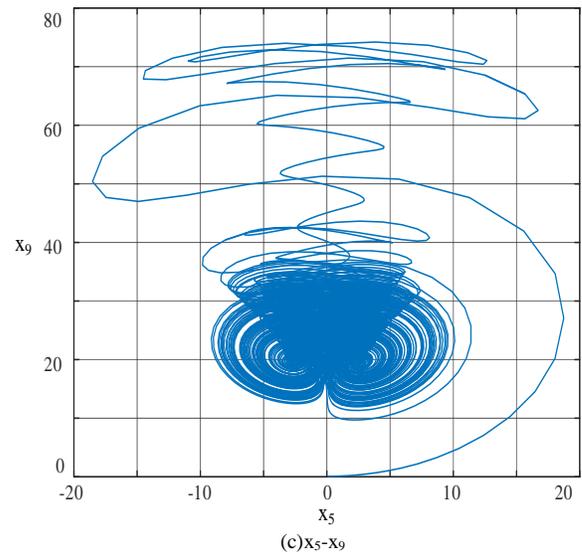

(c)$x_5$-$x_9$

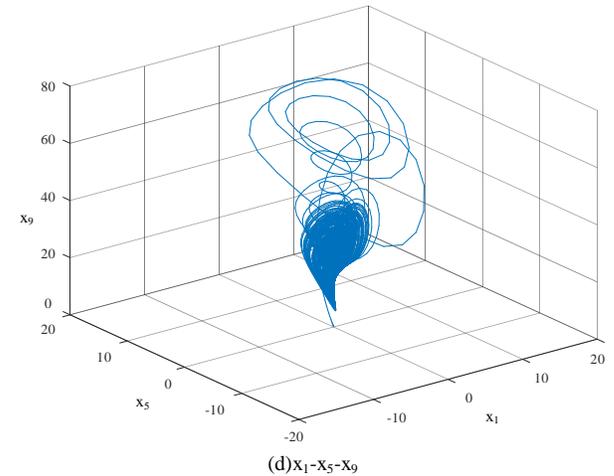

(d)$x_1$-$x_5$-$x_9$

Fig . 2. Phase portraits of system (3)

According to [21], Lyapunov exponent is a numerical characteristic representing the average exponential divergence rate of adjacent trajectories in phase space. It is an important index to analyse the dynamic characteristics. Only when there are positive, negative values and zero, the system is chaotic. The Lyapunov exponents of system (3) are LE1= 2.028, LE2 = 1.913, LE3 =1.863, LE4 =1.502, LE5 = 0.000, LE6 =-0.316, LE7 = -5.380, LE8 = -5.524, LE9 = -5.809. The corresponding Lyapunov exponents are depicted in Fig. 3. The Lyapunov exponent of the proposed 9-dimensionol system is (+, +, +, +, 0, -, -, -, -). Therefore, the system is chaotic.

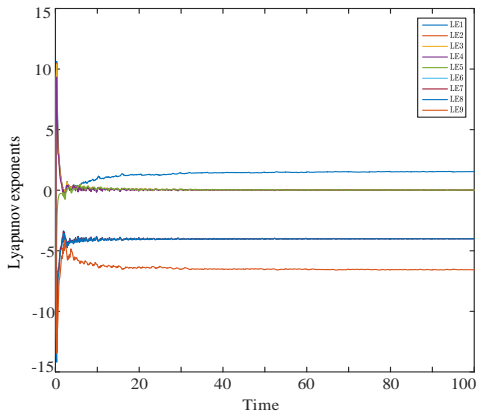

Fig . 3.  The Lyapunov exponent spectrum of system (3)

### B.  Bifurcation Diagram

With different system parameters values, the dynamic motion state of the system will change. This phenomenon is called bifurcation [22]. When there is only one point in the bifurcation diagram, it shows that the system is in a steady state within this parameter range. On the contrary, when there are countless points in the bifurcation diagram, it means that the system is chaotic within this parameter range.  Therefore, the bifurcation diagram varying with parameters can be used to analyze the performance characteristics of the system.

The bifurcation diagram of variable $x_1$ varying with parameter $a$ is shown in Fig. 4. According to the Fig. 4, the system ceaselessly branches between diverse states with the change of $a$ , and the system comes to a chaotic state at last.

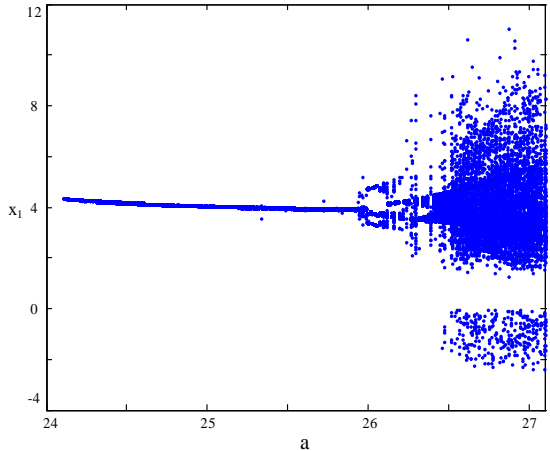

Fig . 4.  The bifurcation of system (3)

### C.  0-1 test

By calculating the transformation variables of the sequence $s(n)$ and $p(n)$ , the system state is judged. This method is called 0-1 test.

$$s(n) = \sum_{j=1}^{n} x(j)\sin(jc), n = 1, 2, ..., N \qquad (4)$$

$$p(n) = \sum_{j=1}^{n} x(j)\cos(jc), n = 1, 2, ..., N \qquad (5)$$

where $c \in (0, \pi)$ , $x(j)(j=1,2,...,N)$ is the test sequence.

Next, verify the trajectory of $p(n) - s(n)$ . If the trajectory shows the Brownian motion, the test sequence is chaotic. The "0-1 test" diagram of system (3) is shown in Fig. 5. It can be seen that the system (3) shows the Brownian motion. Therefore, it is chaotic.

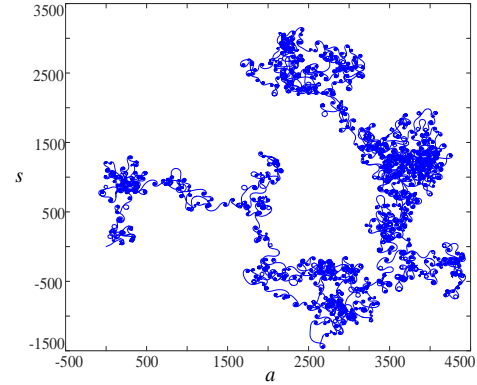

Fig . 5.  0-1 test diagram

### D.  Complexity Analysis

The *SE* algorithm and $C_0$ algorithm are used to verify and analyze the complexity of chaotic systems when the two parameters change. The *SE* algorithm obtains the spectral entropy from the Shannon entropy algorithm. The $C_0$ algorithm divides the sequence into regular and irregular parts. Then it calculates the proportion of irregular parts in the whole sequence. The chromatogram of parameter $a$ varying with parameter $c$ is shown in Fig. 6. The darker the color is, the higher the complexity is.

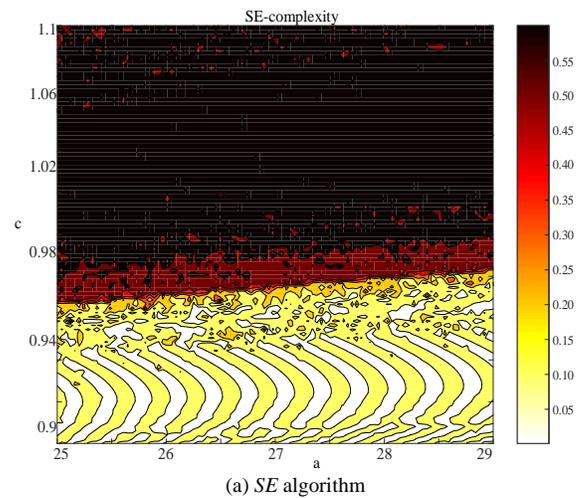

(a) *SE* algorithm

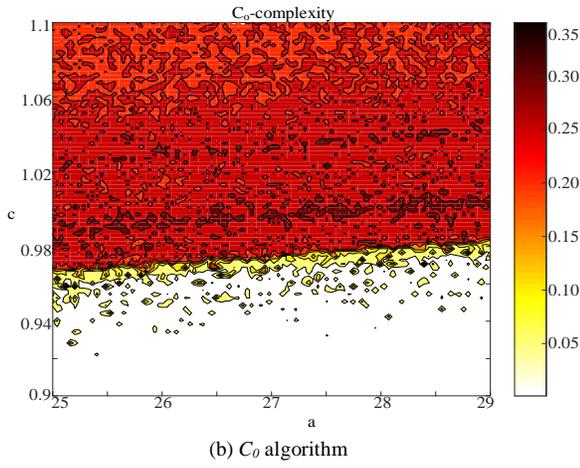

Fig . 6. The Chromatogram of $x_1$ sequence

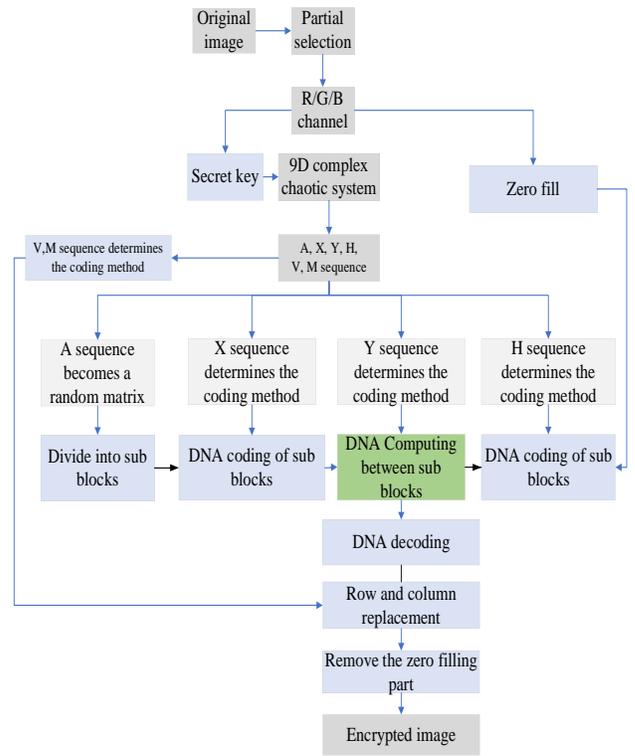

Fig. 7. The flow chart of image encryption algorithm

## III. Encryption Algorithm And Its Comparison

Generally speaking, encrypted data will not be partially selected to ensure the integrity of data. However, when the equipment network breaks down, the real-time reporting of fault data and ensuring the safe transmission of data are equally important. The flow chart of the proposed encryption algorithm is shown in Fig. 7. The steps of the image encryption algorithm are as follows:

**Step 1**: Partially selected image. And separate the image channel and change it into R, G and B channels. Then fill the remaining part in zero pixel value.

**Step 2**: Select the gray value of each channel and its combination as the key. And the image is divided into $4 \times 4$ subblocks. Call it A1.

**Step 3**: Randomly select six sequences from the system (3). Then name them A, X, Y, H, V, M sequences. Change the A sequence into a matrix of the same image size. And it is divided into $4 \times 4$ subblocks. Name it A2.

**Step 4**: Encode A1 and A2 using the DNA coding scheme. The coding rule is determined by the X and Y sequences respectively. Name them B1 and B2.

**Step 5**: Encode B1 and B2 using the DNA computing scheme. The computing scheme is determined by the H sequence. Call it C1.

**Step 6**: Decoding C1 using the DNA coding scheme. And replace the row and column. The row and column replacement orders are determined by V and M sequences respectively. Finally, the zero filling part is removed.

Set $a = 27$, $b = 23$, $c = 1$ in system (3) and initial condition is the pixel average value of original image. In this encryption algorithm, the "Lena" image is an encrypted object. The encryption process is depicted in Fig. 8, where Fig. 8(a) is the original image, Fig. 8(b) is the encrypted image, Fig.8(c) is the decrypted image and Fig. 8(d) is the partial image. The original features and information are masked in Fig. 8(e).

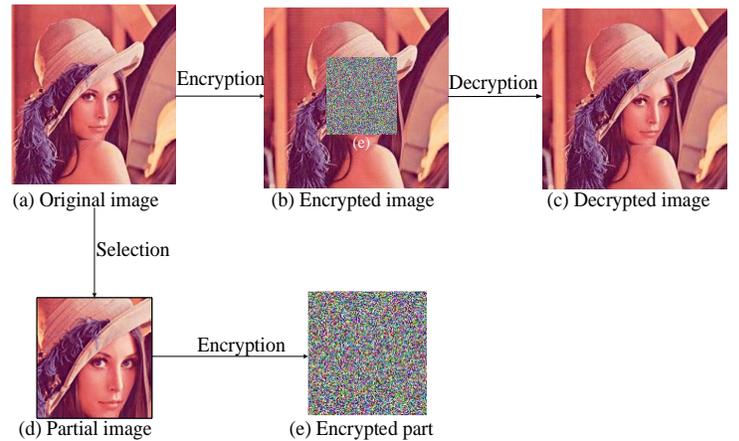

Fig. 8. The flow chart of experimental results

### A. Reconstruction quality analysis

In order to measure the distortion of the decrypted image, structural similarity (SSIM) is introduced. SSIM contains three independent components: Luminance, Contrast and Structure. The SSIM can be expressed as follows:

$$
\begin{cases}
U(X,Y) = \dfrac{2k_X k_Y + D_1}{k_X^2 + k_Y^2 + D_1} \\[2mm]
O(X,Y) = \dfrac{2n_X n_Y + D_2}{n_X^2 + n_Y^2 + D_2} \\[2mm]
T(X,Y) = \dfrac{n_{XY} + D_3}{n_X n_Y + D_3} \\[2mm]
SSIM = U(X,Y)O(X,Y)T(X,Y)
\end{cases}
\tag{6}
$$

where $k_x$ and $k_y$ are the mean of images $X$ and $Y$. $n_x$ and $n_y$ are the variance of images $X$ and $Y$. $n_{xy}$ is the covariance of images $X$ and $Y$. $D_1$ and $D_2$ are two constants with small value when the denominator is up to zero. $D_3 = \frac{1}{2}D_1$. $U(X,Y)$ is the luminance, $O(X,Y)$ denotes contrast, $T(X,Y)$ represents structure. The SSIM of Fig. 8(a) and (c) is 1 by calculation.

Next, subtract the pixel value of the original image from each pixel value of the decrypted image. The result of this calculation is zero, which indicates the image pixel value is restored to normal.

### B. Correlation coefficient

In the horizontal, vertical and diagonal directions, the adjacent pixel values of plaintext and ciphertext images are randomly selected. And calculate the correlation coefficient between two adjacent pixels. The numerical range of correlation coefficient of adjacent pixels is [-1, 1]. If the value is close to 1, the adjacent pixels are strong related. Contrarily, the adjacent pixels are weak related. The formula of the correlation coefficient $r_{uv}$ between $u_i$ and $v_i$ is shown as follows:

$$
E(u) = \frac{1}{N}\sum_{i=1}^{N} u_i
\tag{7}
$$

$$
D(u) = \frac{1}{N}\sum_{i=1}^{N}(u_i - E(u))^2
\tag{8}
$$

$$
Cov(u,v) = \frac{1}{N}\sum_{i=1}^{N}(u_i - E(u))(v_i - E(v))
\tag{9}
$$

$$
r_{uv} = \frac{Cov(u,v)}{\sqrt{D(u)} * \sqrt{D(v)}}
\tag{10}
$$

where $E$ indicates the average pixel value, $D$ and $Cov$ represent the variance and the covariance of pixels, respectively. $r_{uv}$ is the correlation coefficient. $N$ is the number of pixels. $u_i, v_i (i=1,...,N)$ are pixels. As shown in Table I and II, although compared with other algorithms, the confidentiality is slightly inferior, this algorithm has an advantage in shortening the encryption and decryption time.

TABLE I
ADJACENT PIXELS CORRELATION COMPARISON OF DIFFERENT ENCRYPTION

| Cipher Image | H | V | D |
|---|---|---|---|
| Ref. [23] | +0.0083 | -0.0054 | -0.0010 |
| Ref. [24] | -0.0168 | +0.0445 | -0.0022 |
| Ref. [25] | -0.0024 | +0.0035 | +0.0014 |
| Ref. [26] | +0.0010 | -0.0031 | -0.008 |
| Proposed method | -0.0112 | +0.0326 | -0.0032 |

where H is horizontal, V is vertical and D is diagonal.

TABLE II
ENCRYPTION TIME AND DECRYPTION TIME

| Cipher Image | Encryption time | Decryption time |
|---|---|---|
| Ref. [23] | 0.669s | 0.659s |
| Ref. [24] | 2.796s | 2.788s |
| Ref. [25] | 1.840s | 1.827s |
| Ref. [26] | 0.670s | 0.661s |
| Proposed method | 0.170s | 0.177s |

### C. Histogram

The frequency of all gray values can be intuitively seen from the histogram. The pixel distribution of each pixel level can be seen in Fig.9. Fig.9(a), Fig.9(b) and Fig.9(c) are the histogram of plaintext image. And Fig.9(d), Fig.9(e) and Fig.9(f) are the histogram of ciphertext image. The pixel distribution of the plaintext image is uneven. After encryption, the histogram of ciphertext image is almost uniform and is quietly different from those of plaintext image. The results verify that the proposed encryption algorithm is resistant to statistical attacks.

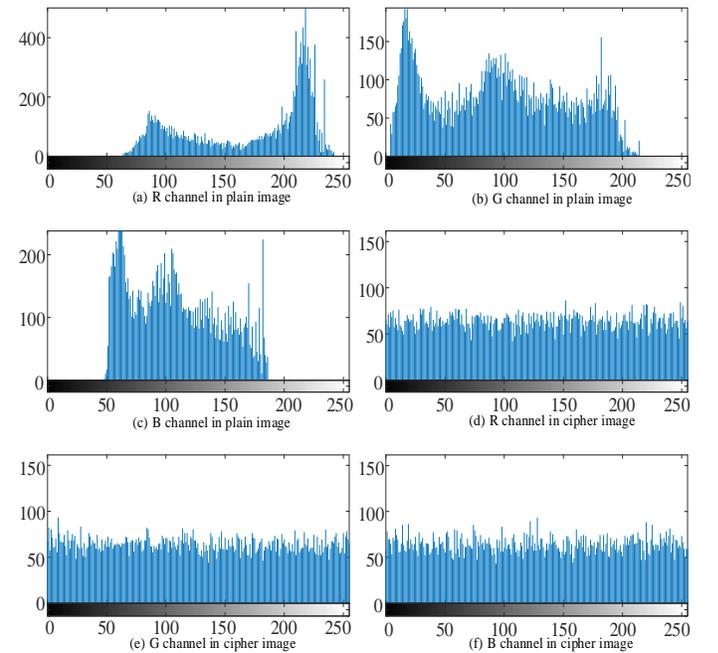

Fig. 9. Histogram

### D. Information Entropy

As a quantitative standard, information entropy is the relative complexity of the image information. The image information entropy with average gray value distribution is relatively close to 8. The formula $F$ is shown as follows:

$$F = -\sum_{i=1}^{T} e(x_i) \log_2 e(x_i) \qquad (11)$$

where $T$ is the maximum gray value 255, and $e(x_i)$ is the gray value probability. As shown in Table III, the information entropy of the proposed algorithm is slightly smaller, but the encryption time is greatly shortened. It means that this algorithm can effectively shorten the encryption time.

TABLE III
THE INFORMATION ENTROPY OF IMAGE

| Image | Plaintext image | Ciphertext image |
|---|---|---|
| Ref. [23] | 7.4375 | 7.9972 |
| Ref. [24] | 7.4375 | 7.9994 |
| Ref. [25] | 7.4375 | 7.9972 |
| Ref. [26] | 7.4375 | 7.9994 |
| Proposed method | 7.4375 | 7.9957 |

## IV. THE ENCRYPTION SCHEME IN THE SMART GRID

Two actual pictures of the distribution network are selected as the encryption objects. Their sizes are respectively $256 \times 256$ and $512 \times 512$. Call them "image 1" and "image 2". The encryption process of "image 1" is shown in Fig. 10, where Fig. 10(a) is the original "image 1", Fig. 10(b) is the encrypted "image 1", Fig. 10(c) is the decrypted "image 1", and Fig. 10(d) is the partial "image 1". The original features and information are masked in Fig. 10(e). The Fig. 11 shows the same encryption process of "image 2".

### A. Image Encryption

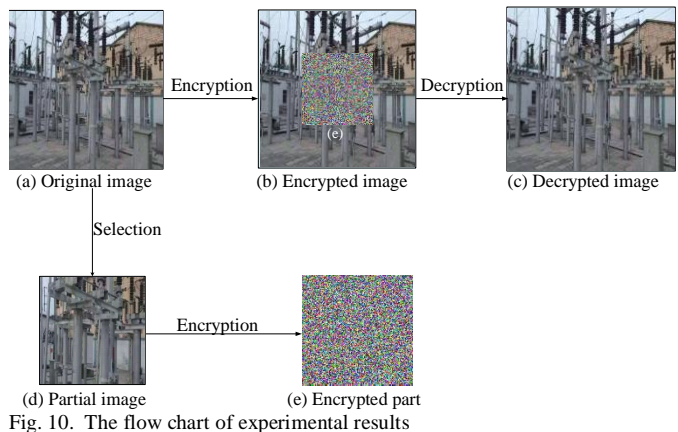

(a) Original image    (b) Encrypted image    (c) Decrypted image

Selection

(d) Partial image    (e) Encrypted part

Fig. 10. The flow chart of experimental results

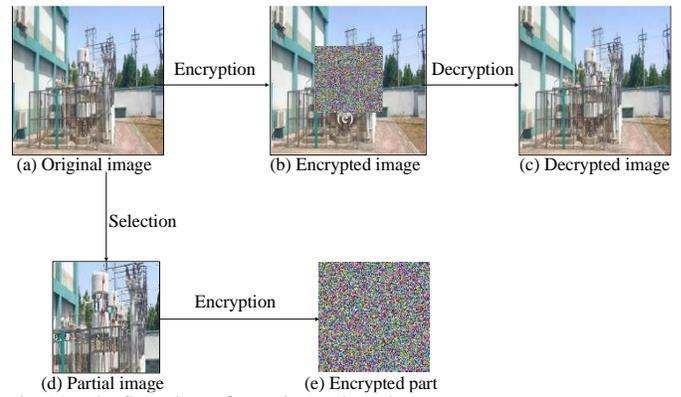

(a) Original image    (b) Encrypted image    (c) Decrypted image

Selection

(d) Partial image    (e) Encrypted part

Fig. 11. The flow chart of experimental results

Then, the encrypted image is analyzed by the histogram, correlation coefficient, information entropy, sensitivity of key and reconstruction quality.

### 1) Histogram

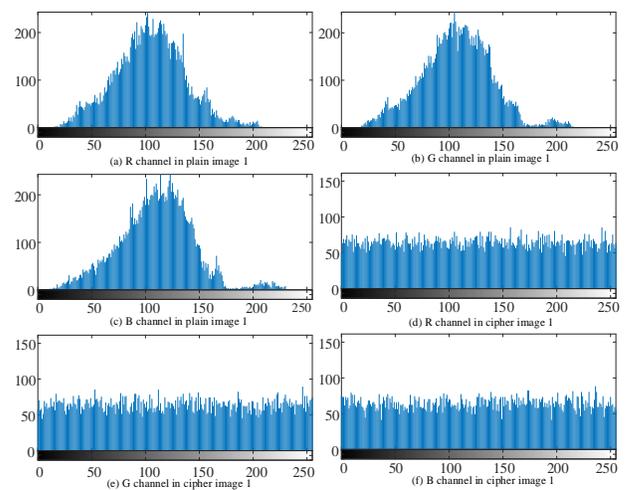

Fig. 12. Histogram of the original "image 1" and the encrypted "image1"

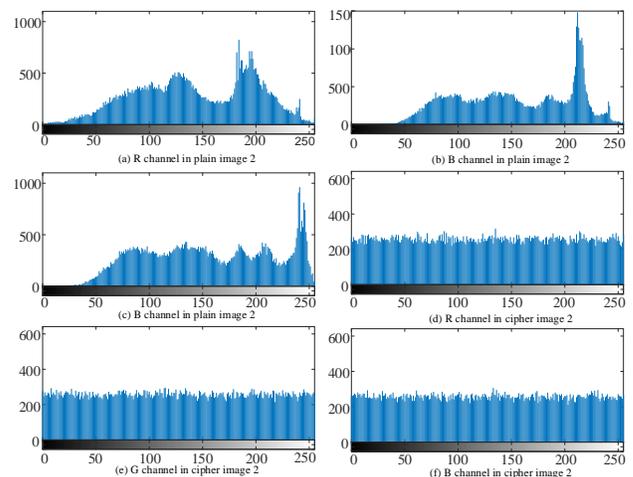

Fig. 13. Histogram of the original "image 2" and the encrypted "image2"

According to the Fig. 12 and Fig. 13, the histogram after encryption is evenly distributed, which can effectively resist statistical attacks.

### 2) Correlation Coefficient

TABLE IV
ADJACENT PIXELS CORRELATION COMPARISON

| Image | H | V | D |
|---|---|---|---|
| Image1 | 0.9053 | 0.7714 | 0.7060 |
| Image2 | 0.9277 | 0.8423 | 0.8061 |
| Cipher image1 | -0.0104 | 0.0011 | 0.0044 |
| Cipher image2 | 0.0133 | -0.0042 | -0.0137 |

According to the Table IV and V, the strong correlation between the pixels of the original image after a short time of encryption is weakened.

TABLE V
ENCRYPTION TIME AND DECRYPTION TIME

| Cipher Image | Encryption time | Dncryption time |
|---|---|---|
| Image 1 | 0.188s | 0.226s |
| Image 2 | 0.611s | 0.816s |

### 3) Information Entropy

TABLE VI
THE INFORMATION ENTROPY OF IMAGE

| Image | Plaintext image | Ciphertext image |
|---|---|---|
| Image 1 | 7.0589 | 7.9964 |
| Image 2 | 7.6724 | 7.9989 |

According to the Table VI, the information entropy of encrypted images with a short encryption time is close to the maximum. It means that this encryption algorithm takes into account both confidentiality and real-time.

### 4) Sensitivity of key

For initial conditions (0.10002, 0.10002, 0.10002, 0.10002, 0.10002, 0.10002, 0.10002, 0.10002, 0.10002), the value of other parameters in system (3) remain unchanged. The generated sequences are used to decrypt the cipher image 1 and 2. According to the Fig .14, these sequences fail to decrypt the cipher image 1 and 2. The proposed encryption algorithm is sensitive to the key.

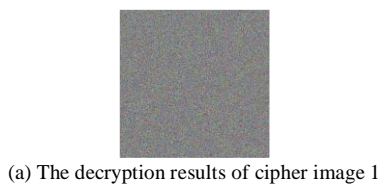

(a) The decryption results of cipher image 1

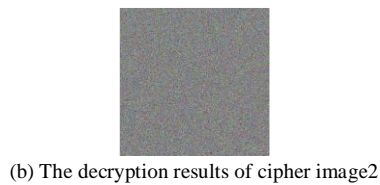

(b) The decryption results of cipher image2

Fig. 14. The Decryption results of wrong key

### 5) Reconstruction quality analysis

TABLE VII
RECONSTRUCTION QUALITY ANALYSIS

| Image | pixel-value difference | SSIM |
|---|---|---|
| Fig. 10(b) and (d) | 0 | 1 |
| Fig. 11(b) and (d) | 0 | 1 |

According to the Table VII, the reconstruction quality of "image 1" and "image 2" is excellent.

From the correlation coefficient, histogram, information entropy, sensitivity of key and reconstruction quality, the encryption algorithm proposed can give consideration to confidentiality and real-time in the image transmission of the smart grid.

### B. Data Encryption

Modbus protocol is adopted in the process of transmitting monitor data. The composition of Modbus protocol is as follows:

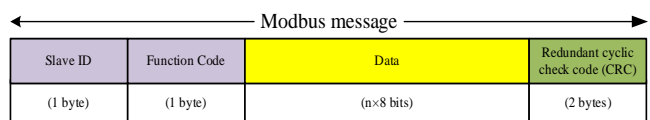

Fig. 15. Modbus RTU mode message frame composition

The Redundant cyclic check code (CRC) is calculated by the transmitting device and placed at the end of the transmitted information frame. The receiving device recalculates the CRC of the received information and compares whether the calculated CRC is consistent with the received CRC. If they are inconsistent, it is considered that the data is abnormal. Therefore, CRC is selected for encryption in this paper. The flow chart of the data encryption algorithm is shown in Fig. 16. This encryption step is similar to the above image encryption step.

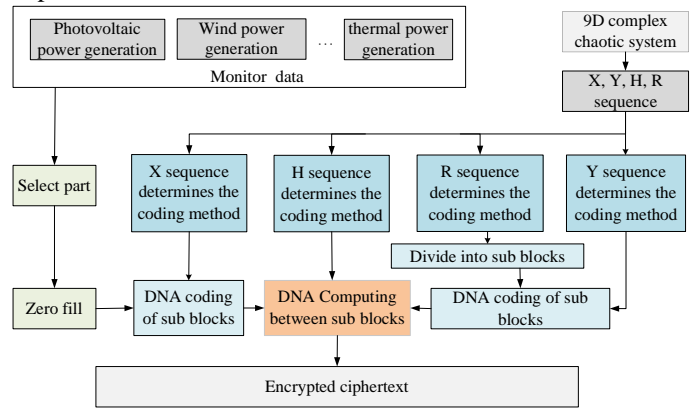

Fig. 16. The flow chart of data encryption algorithm

Next, the randomness of the sequence is analyzed by the NIST test. It is a recognized standard in the encryption field that can be used to evaluate the performance of pseudo-random sequences. And most scholars are convinced that it is a very common and effective method to test pseudo-random sequences.

For convenience's sake, 10, 000, 000 real numbers, generated by system (3), are converted to binary sequences. And adopt them directly as the experimental data of the NIST test suit. If every single item in the NIST test exceeds 0.01 (maximum is 1), it indicates that this test is passed. At the same time, the larger the value is, the stronger the random characteristic of the test sequence is.

As shown in Table VIII, The chaotic sequence generated by system (3) has good random performance and can effectively cover up the information of encrypted data.

TABLE VIII
NIST TEST

| Test category | Value | Pass the test or not |
| --- | --- | --- |
| Approximate Entropy | 0.534146 | pass |
| Block Frequency | 0.350485 | pass |
| Cumulative Sums | 0.534146 | pass |
| FFT | 0.350485 | pass |
| Frequency | 0.911413 | pass |
| Linear Complexity | 0.213309 | pass |
| Longest Run | 0.739918 | pass |
| NonOverlapping Template | 0.911413 | pass |
| Overlapping Template | 0.066882 | pass |
| Random Excursions | 0.213309 | pass |
| Random Excurions Variant | 0.122325 | pass |
| Rank | 0.739918 | pass |
| Runs | 0.739918 | pass |
| Serial | 0.739918 | pass |
| Uiversal | 0.534146 | pass |

## V. CONCLUSIONS

A new 9D complex chaotic system with quaternion is proposed in this paper. Firstly, it is derived from the complex Chen system and quaternion. Secondly, The expansion of variables from real field to complex field is realized. To analyze the performances of the new chaotic system, Lyapunov exponent, phase diagrams, bifurcation diagram, 0-1 test and complexity are proposed. Finally, with the DNA code, an encryption algorithm is proposed based on the system (3). The transmitted images and data are encrypted in the verification experiments.

To analyze the image encryption, the histogram, correlation coefficient, information entropy, sensitivity of key and reconstruction quality are introduced. The data encryption is analyzed by the NIST test. The experiment results show that the proposed algorithm ensures real-time performance on the basis of confidentiality.